\documentclass[twocolumn,amsmath,amssymb,superscriptaddress,prl,a4paper]{revtex4-1}
\usepackage{graphicx}
\usepackage[paperwidth=212mm,paperheight=297mm,centering,hmargin=1.65cm,vmargin=2.6cm]{geometry}

\usepackage{graphicx}                       
\usepackage{color}                          
\usepackage{hyperref}                       
\usepackage{dsfont}
\usepackage{mathrsfs}
\usepackage{bigints}
\usepackage{bbold}
\usepackage{amsthm}
\usepackage{setspace}
\usepackage{epstopdf}

\newcommand{\ket}[1]{\vert#1\rangle}
\newcommand{\bra}[1]{\langle#1\vert}
\newcommand{\outter}[1]{\ket{#1}\bra{#1}}

\newcommand{\beq}{\begin{eqnarray}}
\newcommand{\eeq}{\end{eqnarray}}

\newcommand{\re}{\operatorname{Re}}

\newcommand{\vk}{\mathbf{k}}

\newcommand{\vecr}{\mathbf{r}}

\DeclareGraphicsExtensions{.jpg,.mps,.pdf,.png,.gif}

\begin{document}

\preprint{APS/123-QED}

\author{Antoine Reigue}

\affiliation{Sorbonne Universit\'es, UPMC Univ Paris 06, CNRS UMR 7588, Institut des NanoSciences de Paris, F-75005, Paris, France}

\author{Jake Iles-Smith}
\email{jakeilessmith@gmail.com}
\affiliation{Department of Photonics Engineering, DTU Fotonik, {\O}rsteds Plads, 2800 Kongens Lyngby, Denmark}

\author{Fabian Lux}

\affiliation{Sorbonne Universit\'es, UPMC Univ Paris 06, CNRS UMR 7588, Institut des NanoSciences de Paris, F-75005, Paris, France}

\author{L\'eonard Monniello}

\affiliation{Sorbonne Universit\'es, UPMC Univ Paris 06, CNRS UMR 7588, Institut des NanoSciences de Paris, F-75005, Paris, France}

\author{Mathieu Bernard}

\affiliation{Sorbonne Universit\'es, UPMC Univ Paris 06, CNRS UMR 7588, Institut des NanoSciences de Paris, F-75005, Paris, France}

\author{Florent Margaillan}

\affiliation{Sorbonne Universit\'es, UPMC Univ Paris 06, CNRS UMR 7588, Institut des NanoSciences de Paris, F-75005, Paris, France}

\author{Aristide Lemaitre}

\affiliation{Centre de Nanosciences et de Nanotechnologies, CNRS, Univ. Paris-Sud, Universit\'e Paris-Saclay, C2N - Marcoussis, 91460 Marcoussis, France}

\author{Anthony Martinez}

\affiliation{Centre de Nanosciences et de Nanotechnologies, CNRS, Univ. Paris-Sud, Universit\'e Paris-Saclay, C2N - Marcoussis, 91460 Marcoussis, France}

\author{Dara P. S. McCutcheon}

\affiliation{Quantum Engineering Technology Labs, H. H. Wills Physics Laboratory and Department of Electrical and Electronic Engineering,
University of Bristol, Merchant Venturers Building, Woodland Road, Bristol BS8 1FD, UK}

\author{Jesper M{\o}rk}

\affiliation{Department of Photonics Engineering, DTU Fotonik, {\O}rsteds Plads, 2800 Kongens Lyngby, Denmark}

\author{Richard Hostein}

\affiliation{Sorbonne Universit\'es, UPMC Univ Paris 06, CNRS UMR 7588, Institut des NanoSciences de Paris, F-75005, Paris, France}

\author{Valia Voliotis}
\email{voliotis@insp.jussieu.fr}

\affiliation{Sorbonne Universit\'es, UPMC Univ Paris 06, CNRS UMR 7588, Institut des NanoSciences de Paris, F-75005, Paris, France}

\title{Probing electron-phonon interaction through two-photon interference in resonantly driven semiconductor quantum dots}

\begin{abstract}
We investigate the temperature dependence of photon coherence properties through two photon interference (TPI) measurements from a single QD under resonant excitation. 
We show that the loss of indistinguishability is only related to the electron-phonon coupling without being affected by spectral diffusion.
Through these measurements, and a complementary microscopic theory, we identify two independent separate decoherence processes each associated to phonons. 
Below $10 \, \mathrm{K}$, we find that the relaxation of the vibrational lattice is the dominant contribution to the loss of TPI visibility. This process is non-Markovian in nature, and corresponds to real phonon transitions resulting in a broad phonon sideband in the QD emission spectra. 
Above $10 \, \mathrm{K}$, virtual phonon transitions to higher lying excited states in the QD become the dominant dephasing mechanism, this leads to broadening of the zero phonon line, and a corresponding rapid decay in the visibility. The microscopic theory we develop provides analytic expressions for the dephasing rates for both virtual phonon scattering and non-Markovian lattice relaxation.
\end{abstract}

\maketitle

Many recent developments in quantum information processing rely on the use of solid-state qubits that can emit indistinguishable single photons on demand~\cite{PhysRevA.69.032305, 0034-4885-75-12-126503}. However, 
maintaining the coherence between consecutively emitted photons remains a true challenge for realising a deterministic source of identical photons. A promising candidate for the development of such a source 
are self-assembled semiconductor quantum dots (QDs) embedded in photonic nanostructures~\cite{RevModPhys.87.347, Gao:2015aa}. 

{Despite impressive milestones in the development of these devices,} a QD naturally couples strongly to its surrounding solid-state matrix, constituting an inherently open quantum system. The excitonic degrees of freedom are heavily influenced by the vibrational modes~\cite{{ramsay2010damping,ramsay2010phonon},McCutcheon2013,nazir2016modelling}, fluctuating charges~\cite{houel2012probing}, and nuclear spins~\cite{kuhlmann2013charge,*Kuhlmann2015} of the host material, all of which lead to dephasing, therefore suppressing the coherence properties of the emitted photons.
Decoherence may be reduced by enhancing the emission rate through the Purcell effect, or using resonant excitation to minimise laser-induced dephasing, leading to bright single-photon sources with near unity indistinguishability~\cite{gazzano2013bright, ding2016demand, he2013demand, somaschi2016near, PhysRevB.90.041303}.

Still, there are a number of open questions regarding the role of phonon processes on the coherence properties of emitted photons.
Recent experimental and theoretical work has demonstrated the importance of a microscopic model to understand the role of phonons on the emission properties of QDs~\cite{mccutcheon2010quantum,nazir2016modelling, glassl2012interaction,roy2011phonon, kaer2012microscopic, majumdar2011phonon, wei2014temperature, wilson2002quantum,hohenester2010cavity,iles2016fundamental,Iles-Smith:16,ilessmithFuture}.
Examples include excitation-induced dephasing of excitonic Rabi oscillations \cite{ramsay2010damping, monniello2013excitation}; sideband linewidth in resonance fluorescence (RF)~\cite{flagg2009resonantly, ulrich2011dephasing, wei2014temperature};  and temperature-dependent Rabi frequency renormalisation \cite{ramsay2010phonon, wei2014temperature}.
In the examples given, the excitonic degrees of the QD are assumed to couple linearly to the phonon environment, inducing thermalisation in the QD eigenbasis~\cite{McCutcheon2013,nazir2016modelling}, leading to a broad non-Markovian sideband in the QD emission spectra~\cite{{kaer2013role, *kaer2014decoherence}}.  

Here, we present a combined experimental and theoretical investigation of the coherence properties of photons emitted by a quantum dot, allowing us, unambiguously, to separate real and virtual de-coherence processes due to phonons and their temperature dependencies.
To do so, we take temperature dependent measurements of two-photon interference (TPI) in a Hong-Ou-Mandel (HOM) configuration using strictly resonant excitation conditions. 
{We show that TPI measurements are not affected by spectral diffusion due to fluctuating charges, as this process is slow compared to the emission time interval between the two interfering photons. Therefore loss of indistinguishability is attributed only to electron-phonon scattering. 
Through temperature dependent TPI measurements, we demonstrate that linear electron-phonon coupling is not sufficient to capture the trend in TPI visibility.}
In order to describe the behaviour observed, we develop a microscopic model, based on polaron theory, to include phonon induced virtual transitions to higher lying states in the QD~\cite{muljarov2004dephasing,grange2009decoherence}; this leads to temperature dependent broadening of the Zero Phonon Line (ZPL).
Our formalism, allows us to derive analytic forms for the dephasing due to both virtual phonon processes and non-Markovian lattice relaxation. This provides novel insights into the dephasing mechanisms relevant to QD based single photon sources.

\begin{figure}[t!]
\begin{center}
\includegraphics[width=7cm,height=5cm]{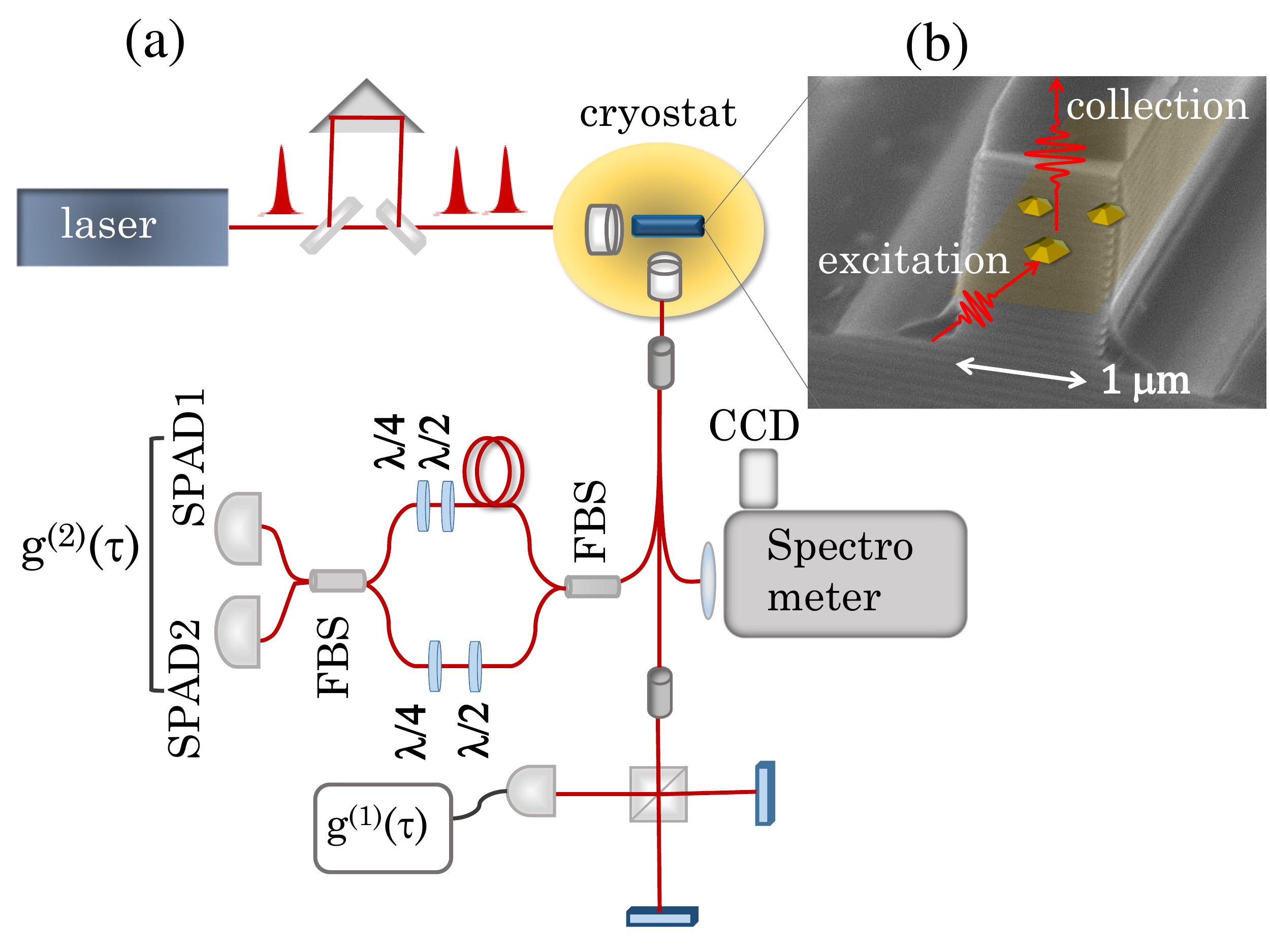}
\end{center}
\caption{(a) Scheme of the experimental set-up. A tunable Ti:sapphire (82 MHz) laser delivers 3-ps pulses and, for HOM experiments, pairs of pulses separated by 3 ns. The laser is focused by a microscope objective on the cleaved edge of one ridge and the RF is collected from the top surface by a second microscope objective. The sample and the two objectives are inside a closed-cycle He temperature-variable cryostat. The signal is coupled to a fibered set-up for either standard spectroscopy, or Michelson interferometry ($g^{(1)}$) or TPI experiments ($g^{(2)}$) using a $3 \, \mathrm{ns}$ unbalanced Mach-Zenhder interferometer. (b) Scanning electron microscopy image of one ridge with the dots schematically drawn in the layer.}
\label{exp_set_up}
\end{figure}

\textit{Experiment} - In contrast to previous temperature dependent TPI measurements~\cite{thoma2016exploring}, our geometry (see Fig.~\ref{exp_set_up})~\cite{melet2008resonant} allows us to use strictly resonant ($s$-shell) excitation removing dephasing due to {relaxation from higher excited states and} time-jitter~\cite{unsleber2015two}. 
Furthermore, the only filtering used in our measurements is due to a low-Q cavity which increases the collection efficiency.  
%
We present results on three different self assembled InAs/GaAs QDs from different samples (labelled QD1, QD2 and QD3 hereafter) excited with resonant $\pi$-pulses. Depending on the QD under study, an additional very weak (few nW) He-Ne laser is added to enhance or recover the RF as reported previously~\cite{nguyen2012optically,PhysRevB.90.041303, nguyen2013photoneutralization}. Experimental details can be found in the Supplemental Material~\cite{supmat}.

The coherence time $T_2$, corresponding to the width of the RF line, can be measured by Fourier transform (FT) spectroscopy using a Michelson interferometer. The contrast of the interference fringes is adjusted by a pseudo-Voigt profile, with an inhomogeneous contribution $\eta$ (see~\cite{supmat}), usually attributed to spectral diffusion effects~\cite{berthelot2006unconventional}. In Table~\ref{ftir_table} we give the values of the measured radiative lifetime $T_1$, of $\eta$ and the ratio $T_2 / 2T_1$ for the three QDs. The corresponding FT spectra for the three QDs are given in Fig.~2 of the Supplemental Material~\cite{supmat}.

\begin{table}[b]
\begin{tabular}{c|c|c|c|c|c|c}
   & $T_1 \ (\mathrm{ps})$ & $\eta$ & $T_2 / 2T_1$ & $g^{(2)}_{\mathrm{HBT}}$ & $V_{\mathrm{TPI}}$ & $\tilde{V}$ \\
  \hline
  \hline
  QD 1 & 1100 & 0.45 & 0.35 & 0.12 & 0.79 & 0.33 \\
  QD 2 & 750 & 0.55 & 0.23 & 0.11 & 0.83 & 0.22 \\
  QD 3 & 670 & 0.10 & 0.71 & 0.07 & 0.83 & 0.68 \\
  \hline
  Errors & 2 \% & $\pm 0.1$ & 10 - 15 \% & $\pm 0.02$ & $\pm 0.04$ & 10 - 15 \%
\end{tabular}
\caption{Values of the experimental parameters (see text for definitions) for the three QDs under resonant excitation. $\tilde{V}$ is the expected TPI visibility from $T_1$ and $T_2$ measurements assuming random dephasing processes (see text). \label{ftir_table}}
\end{table}

\begin{figure}[t]
\begin{center}
\includegraphics[scale=0.3]{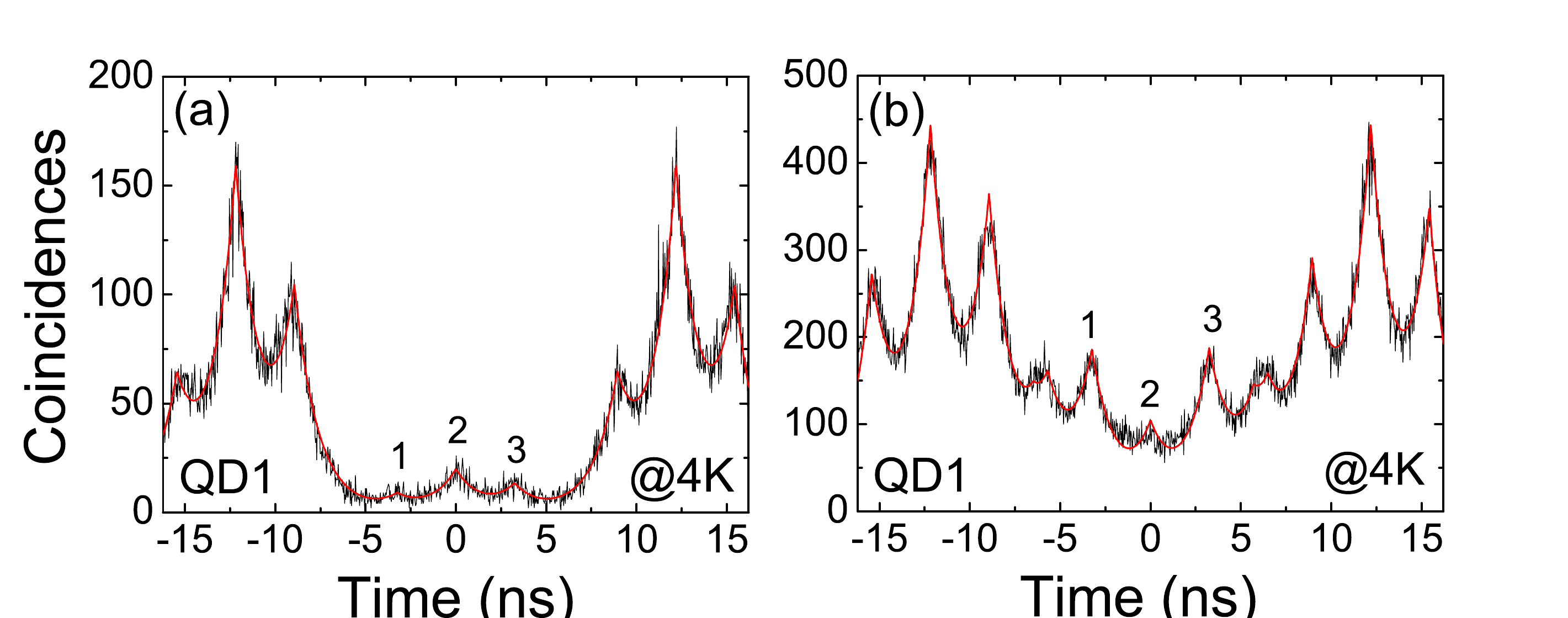} 
\end{center}
\caption{Second-order correlation measurements for QD1 at $ 4 \, \mathrm{K}$ for one hour acquisition time. (a) Coincidences histogram for HBT experiment, we extract $g^{(2)}_{\mathrm{HBT}} = 0.12 \pm 0.02$. (b) Coincidences histogram for TPI experiment. After correction by the remaining laser background, we obtain $V_{\mathrm{TPI}} = 0.79 \pm 0.03$.}
\label{HBT_HOM_maintext}
\end{figure}

Second-order correlation measurements have been performed, allowing characterisation of the single-photon emission purity and  indistinguishability. Fig.~\ref{HBT_HOM_maintext} shows the results obtained for QD1 at $4 \, \mathrm{K}$. Single-photon interferences in a Hanbury-Brown-Twiss (HBT) experiment (Fig.~\ref{HBT_HOM_maintext}a) clearly show an antibunching with a low multiphoton probability $g^{(2)}_{\mathrm{HBT}}$, the values are given for the three QDs in Table~\ref{ftir_table}. We attribute the background correlations at $0$ and $\pm 3 \, \mathrm{ns}$ delay to the remaining scattered laser. The unusual shape of the histogram of coincidences is explained by the fact that the single photons pass through the two arms of the Mach-Zehnder interferometer used for the HOM setup (see Fig.~\ref{exp_set_up}a and~\cite{supmat}). A multi-exponential decay fit (red line) is used to extract the value of $g^{(2)}_{\mathrm{HBT}}$ taking into account the overlapping of the different peaks. Fig.~\ref{HBT_HOM_maintext}b shows the raw histogram of the TPI coincidences for QD1 at $4 \, \mathrm{K}$. The signature of the 
indistinguishability of two successively emitted photons corresponds to the small area of peak 2 compared with peaks 1 and 3. 
The TPI visibility $V_{\mathrm{TPI}}$ (see Table~\ref{ftir_table}) is deduced from the second-order correlation function at zero delay $g^{(2)}_\mathrm{HOM}$ corrected by the remaining scattered laser and by the experimental imperfections (contrast of the Mach-Zehnder interferometer and not perfect 50/50 fibered beam splitters)~\cite{santori2002indistinguishable,supmat}.

In the literature, it is commonly accepted that the visibility of TPI experiments can be obtained from the ratio $T_2/2T_1$, assuming random dephasing processes by phonons and charges~\cite{bylander2003interference}, and labelled $\tilde{V}$ hereafter. The values of $\tilde{V}$ are given  in Table~\ref{ftir_table} and correspond to a poor degree of indistinguishability. This is in striking contrast with the results obtained by TPI experiments where $V_{\mathrm{TPI}} \sim 0.80$ for all QDs at 4~K. The significant difference can be explained by recognising the distinct characteristic timescales of the two kind of experiments which probe different physical dephasing processes
~\cite{gazzano2013bright, ding2016demand, thoma2016exploring, gold2014two, delbecq2016quantum}. Indeed, because of the long acquisition time (seconds) during $T_2$ measurements, the visibility $\tilde{V}$ integrates the interaction processes with the acoustic phonon bath (ps range) and the electrostatic environment ($\mu$s range~\cite{arnold2014cavity}). In contrast, TPI experiments have a characteristic timescale defined by the nanosecond time-delay between pulses, therefore only the QD--acoustic phonon interactions are probed, thus leading to a much higher visibility. This is corroborated by the inhomogeneous contribution $\eta$, which is large when the interaction between the QDs and the charges is dominant~\cite{berthelot2006unconventional}, and corresponds then to low value of $\tilde{V}$. At variance with our results, it has been recently reported that TPI experiments probe both charge fluctuations and phonon-induced dephasing~\cite{gazzano2013bright, thoma2016exploring}. 
However, in these experiments a non-resonant excitation has been used explaining a probable laser-induced dephasing.
Moreover, very recently, Wang~\emph{et al.}~\cite{wang2016near} have also shown by increasing the time-delay between the emission of the two photons, that spectral diffusion has no effect (in the ns range) on the visibility when the QD is resonantly pumped.

\begin{figure}[t]
\begin{center}
\includegraphics[width=0.75\columnwidth]{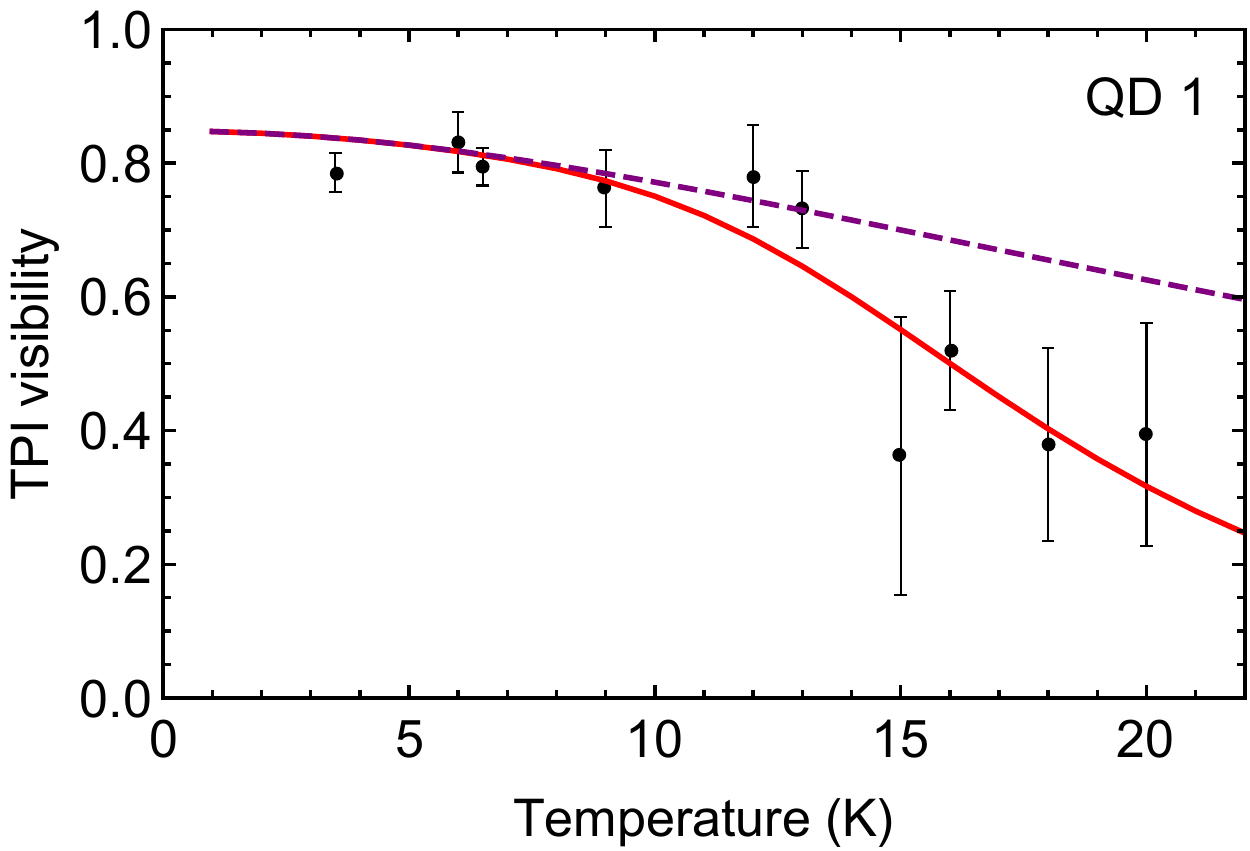}
\end{center}
\caption{Plot showing the visibility of the measured TPI for QD1 as a function of temperature (black points). The data are fitted with Eq.~\ref{ind}; the full expression is used to produced the solid red curve, which fits accurately over the full temperature range. The dashed purple curve, shows only the effect of the phonon sideband on the TPI visibility.
}
\label{V_fct_T}
\end{figure}

The clear separation of timescales described above means that TPI measurements effectively isolate the phonon processes from other dephasing mechanisms. Thus, through temperature dependant TPI experiments performed on a resonantly-driven single QD we can directly asses the importance of phonon processes on the coherence properties of subsequently emitted photons. The measured $V_{\mathrm{TPI}}$ as a function of the temperature for QD1 is presented in Fig.~\ref{V_fct_T} where a clear loss of indistinguishability around $10 \, \mathrm{K}$ is observed. To describe this behaviour below we develop a theoretical model, which fully captures the observed trend in visibility. 
\begin{figure}[t]
\begin{center}
\includegraphics[width=0.75\columnwidth]{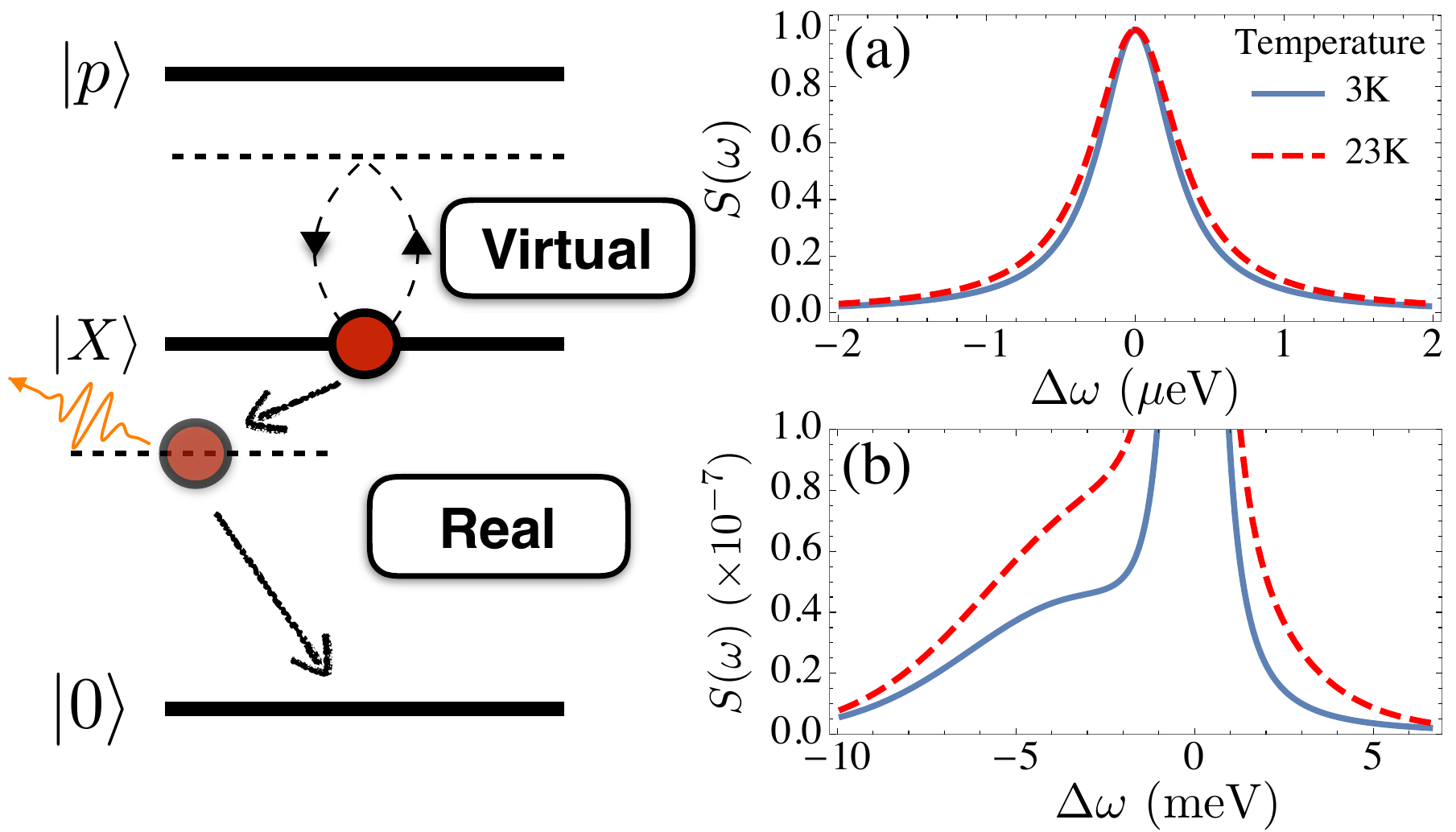} 
\end{center}
\caption{Schematic of real and virtual transition in a QD (left). (a) and (b) show the impact of these transitions to the QD spectra.  Virtual transitions give broadening of the ZPL as in (a). Real transitions lead to a broad phonon sideband illustrated in (b). }
\label{schem}
\end{figure}

\textit{Microscopic model} - 
%
In principle, the many-body electron-phonon interaction Hamiltonian contains all possible electronic configurations of the QD~\cite{mahan2013many}. 
However, when calculating the effect of phonons on the exciton dynamics the energy separation between the $s$- and $p$-orbitals of the QD 
is typically a few {tens} meV, and thus significantly {higher} than the average phonon energies in low temperature experiments. 
With this in mind, any transitions {from the first excited to another electronic state of the QD} must be virtual in nature.
Following Muljarov and Zimmerman~\cite{muljarov2004dephasing} we derive an effective Hamiltonian theory~\cite{cohen1992atom}, treating the charge density operator as a perturbation, with the first- and second-order terms capturing real and virtual phonon transitions respectively, as shown schematically in Fig.~\ref{schem}. 

We start by explicitly considering the ground state $\ket{0}$, and single exciton in the $s$-shell denoted $\ket{X}$, with excitonic splitting $\omega_X$. By using an effective Hamiltonian theory~\cite{cohen1992atom} we can eliminate off diagonal transitions to higher lying phonon states~\cite{muljarov2004dephasing}, yielding the Hamiltonian, $H = H_0 + \ket{X}\bra{X}(\hat{V}+ \hat{V}_Q)$. Here
$
H_0 = \omega_X\ket{X}\bra{X} + \sum_\vk\nu_\vk b_\vk^\dagger b_\vk$ 
is the free Hamiltonian of the QD and phonon environment, where $b_\vk$ is the annihilation operator of a phonon with wave vector $\vk$ and frequency $\nu_\vk$. 
The electron-phonon interaction has two contributions, the first is the standard linear electron-phonon coupling $\hat{V} = \sum_\vk g_\vk(b^\dagger_\vk + b_\vk)$, describing the displacement of the lattice due to the change in charge configuration of the QD~\cite{nazir2016modelling}. 
The coupling strength is quantified through the matrix element $g_\vk = \sum_{a = e,h} M_{a,\vk}^{11}$, where for deformation potential coupling,  
$$
M_{a,\vk}^{ij} = \sqrt{\frac{\nu_\vk}{2\varrho c_s^2\mathcal{V}}}D_a\int d^3r\psi_{i a}^\ast(\vecr)\psi_{j a}(\vecr)e^{i\vk\cdot\vecr},
$$ 
is the matrix element corresponding to phonon induced transition between the $i^{th}$- and $j^{th}$ electronic states. 
Here $\varrho$ is the mass density, $c_s$ is the speed of sound in the material, and $\mathcal{V}$ is the phonon normalisation volume. This matrix element is dependent on the wavefunction $\psi_{i,e/h}(\vecr)$, of the confined electron/hole, and the corresponding deformation potential $D_a$. 

The second term is quadratic in phonon operators~\cite{muljarov2004dephasing} $V_Q =  \sum_{\vk,\vk^\prime} f_{\vk,\vk^\prime}(b^\dagger_\vk + b_\vk)(b^\dagger_{\vk^\prime}+ b_{\vk^\prime})$, and describes virtual phonon transitions between the first exciton state and higher lying excited states.
The effective coupling strength for the quadratic coupling takes the form $f_{\vk,\vk^\prime} = \sum_{a=e,h}\sum_{j>1}M^{1j}_{a,\vk}M^{j1}_{a,\vk^\prime}[\omega^{a}_m - \omega^a_1]^{-1}$, where $\omega^{e/h}_m$ is the energy of the $m^{th}$ electron/hole energy level.


To model the impact of the phonon processes on the photon indistinguishability we make use of the polaron transformation~\cite{mccutcheon2010quantum,roy2011phonon,nazir2016modelling,PhysRevB.92.205406} {through the operator}, $\mathcal{U} = \ket{0}\bra{0} + \ket{X}\bra{X}e^{S}$, where $S = \sum_\vk g_\vk(b^\dagger_\vk -b_\vk)/\nu_\vk$.
This transformation leads to a state dependent displacement of the phonon environment, removing the linear electron phonon coupling.
Applying this transformation to our quadratic Hamiltonian we obtain $H_V = \mathcal{U}^\dagger H\mathcal{U}=(\tilde\omega_X+\hat{V}_Q)\ket{X}\bra{X}+ \sum_\vk\nu_\vk b_\vk^\dagger b_\vk$. 
Notice the residual quadratic electron-phonon coupling, and that the QD resonance is shifted $\tilde\omega_X = \omega_X + \sum_\vk g^2_\vk/\nu_\vk$ . 
From this Hamiltonian we may derive a master equation for the reduced {density matrix} of the QD in the polaron frame, $\chi$, which takes the simple pure dephasing form:
$$
\dot\chi(t) = -i\tilde\omega_X\left[\sigma^\dagger\sigma , \chi(t)\right] +{\Gamma}\mathcal{L}_{\sigma}[\chi(t)]+2{\gamma_{pd}}\mathcal{L}_{\sigma^\dagger\sigma}[\chi(t)],
$$
where $\sigma = \ket{0}\bra{X}$ is the dipole transition operator, and $\mathcal{L}_O[\chi] = O \chi O^\dagger - \left\{O^\dagger O,\chi\right\}/2$. 
Here $\Gamma$ is the {radiative recombination rate} of the QD. If we consider only virtual transitions between the lowest exciton states and the next highest states~\footnote{This is a manifold of three degenerate states with $p$-symmetry.}, then we may find an analytic form for the pure dephasing rate due to the virtual phonon transition~\cite{supmat}:
\begin{equation}
\gamma_{pd} = \frac{\alpha^2 \mu}{\nu_c^4}\int\limits_0^\infty \nu^{10}e^{-2\nu^2/\nu_c^2} n(\nu)(n(\nu) + 1)d\nu,
\end{equation}
where $n(\nu) = [e^{-\beta\nu}-1]^{-1}$.  $\alpha$ and $\mu$  describe the electron-phonon coupling strength and the probability of virtual phonon processes respectively. The cut-off frequency, $\nu_c$, is directly related to the QD confinement length.

To calculate the visibility of two photon interference we must associate the excitonic degrees of freedom in the QD to the emitted field.
 In the polaron frame one obtains the Heisenberg picture field operator $\hat{E} (t)= \sqrt{\Gamma/2\pi}\sigma(t) B_-(t)$, where the standard expression for a dipole emitter is modified by the phonon displacement operator $B_{\pm}(t) = e^{\pm S(t)}$ ~\cite{iles2016fundamental,ilessmithFuture}.
With this expression we obtain the polaron frame first-order correlation function, $g^{(1)}(t,\tau) = \frac{\Gamma}{2\pi}\langle B_+(\tau)B_-\rangle\langle\sigma^\dagger(t+\tau)\sigma(t)\rangle$. 
The second term in this equation describes emission through the ZPL. 
The first term is the phonon correlation function which takes the form $\langle B_+(\tau)B_-\rangle = B^2\exp(\varphi(\tau))$,  where $B = \exp(-\varphi(0)/2)$ is the Frank-Condon factor, and $\varphi(\tau) = \alpha\int_0^\infty \nu\exp(-\nu^2/\nu_c^2)(\coth\left(\beta\nu/2\right)\cos(\nu\tau)-i\sin(\nu\tau))d\nu$.
This function decays on a timescale related to the inverse of the cut-off frequency, which is typically on the order of picoseconds, leading to a the broad phonon sideband in the spectra of the system. A detailed derivation of these expression is given in the supplement.

Following Ref.~\cite{ilessmithFuture} we obtain an analytic form for the indistinguishability including the phonon sideband contribution~\cite{supmat}:
\begin{equation}\label{ind}
\mathcal{I} = \frac{\Gamma}{\Gamma+2\gamma_{pd}}\left(\frac{|h(0)|^2B^2}{|h(0)|^2B^2 + \mathcal{F}(1-B^2)}\right)^2,
\end{equation}
where $|h(0)|^2=(\kappa/2)^2(\delta^2+(\kappa/2)^2)^{-1}$ with $\kappa$ the cavity width and $\delta$ the QD--cavity detuning. 
The first factor gives the contribution of photons emitted through the ZPL, while the second describes the reduction of the indistinguishability from photons emitted through the phonon sideband. 
The temperature dependent factor, $\mathcal{F}$, quantifies the unfiltered fraction of the phonon sideband extracted from the low-Q cavity~\footnote{The value of $\mathcal{F}$ varies between $0.19$ for $T\sim4$~K and $0.33$ for $T\sim22$~K.}.

We use the above expression to fit the experimental data given in Fig.~\ref{V_fct_T}, using a least min squared fitting, we find an 
optimum fit for parameters $\alpha = 0.0082 \, \mathrm{ps}^2$, $\nu_c = 7.9 \, \mathrm{ps}^{-1}$ and $\mu = 4.4 \times 10^{-4} \mathrm{ps}^2$. {$\alpha$ 
and $\mu$ depend only on the material parameters and reasonable agreement is found when compared to the theoretical values. The cut-off frequency 
gives a characteristic confinement length of the order of one nanometre which is the right order of magnitude for typical self assembled QD.} 
The fit captures the qualitative and quantitative behaviour of the data. Furthermore, from the simple form of the expression given in Eq.~\ref{ind}, 
we can analyse the contributions to the indistinguishability due to the phonon sideband, and the virtual transitions to higher lying QD states. 
{Similar results have been obtained on QD2 and are presented in the Supplemental Material~\cite{supmat}.}

{Below $10 \, \mathrm{K}$,} the average energy of phonons $k_BT$ is not sufficient to induce virtual transitions ($\gamma_{pd} \ll \Gamma$). 
If we use the extracted parameters to consider only the sideband contribution (dashed curve) we observe a good fit at low temperatures, with a $\sim10\% $ reduction in the indistinguishability, suggesting that emission via the phonon sideband is the principal cause for the reduction of TPI visibility. However, {above $10 \, \mathrm{K}$} the data demonstrate a rapid decrease in the indistinguishability not captured by the sideband theory.
At these temperatures $k_BT$ is sufficient to induce virtual transitions between the $s-$ and $p-$states of the QD, leading to pure dephasing of the ZPL ($\gamma_{pd} \sim \Gamma$), and consequently a suppression of the indistinguishability. 

The fact that real and virtual phonon processes occur on separate temperature scales is of vital importance to the development of solid-state single photons sources. 
Although the phonon sideband is a persistent problem at low temperatures, it may be easily removed through spectral filtering, or use of a high Q-cavity~\cite{ilessmithFuture}.
Though this could reduce the efficiency of the source, transform limited photons may be obtained as $\gamma_{pd}\approx0$ at low temperature.
In contrast, the broadening of the ZPL cannot be removed through simple filtering, instead one must rely on Purcell enhancement to reduce its influence.
However, the sensitivity of virtual phonon processes to temperature will place significant limitations on the operating regimes of QD systems. 

In summary, through temperature dependent TPI measurements, we have demonstrated 
that both real and virtual phonon transitions, occurring on very different timescales, play a role in reducing the indistinguishability of photons emitted from QDs under resonant excitation. 
Using a rigorous microscopic theory, we provide analytic expressions for the dephasing due to these mechanisms, providing new insights into potential operating regimes of QD single photon sources.
Furthermore, we expect such a general microscopic approach could be used to describe other specific discrete quantum systems coupled to a bosonic reservoir. 


\begin{acknowledgements}
\emph{Acknowledgments} --- We thank R. Grousson , N. Treps, E. Baudin, A. Nazir, P. Tighineanu, and A. S{\o}renson for interesting discussions. 
JIS is supported by Danish Research Council (DFF-4181-00416). 
DPSM acknowledges support from a Marie Sk{\l}odowska-Curie Individual Fellowship (ESPCSS).
This work was partially supported by the French Agence Nationale de la Recherche (ANR-11-BS10-010) and the C’Nano Ile-de-France (No. 11017728).
\end{acknowledgements}

\vspace{-0.35cm}
%

\clearpage
\widetext

\section{Supplementary Material}
\section{Samples and experiment}

\begin{figure}[h!]
\begin{center}
\includegraphics[scale=0.4]{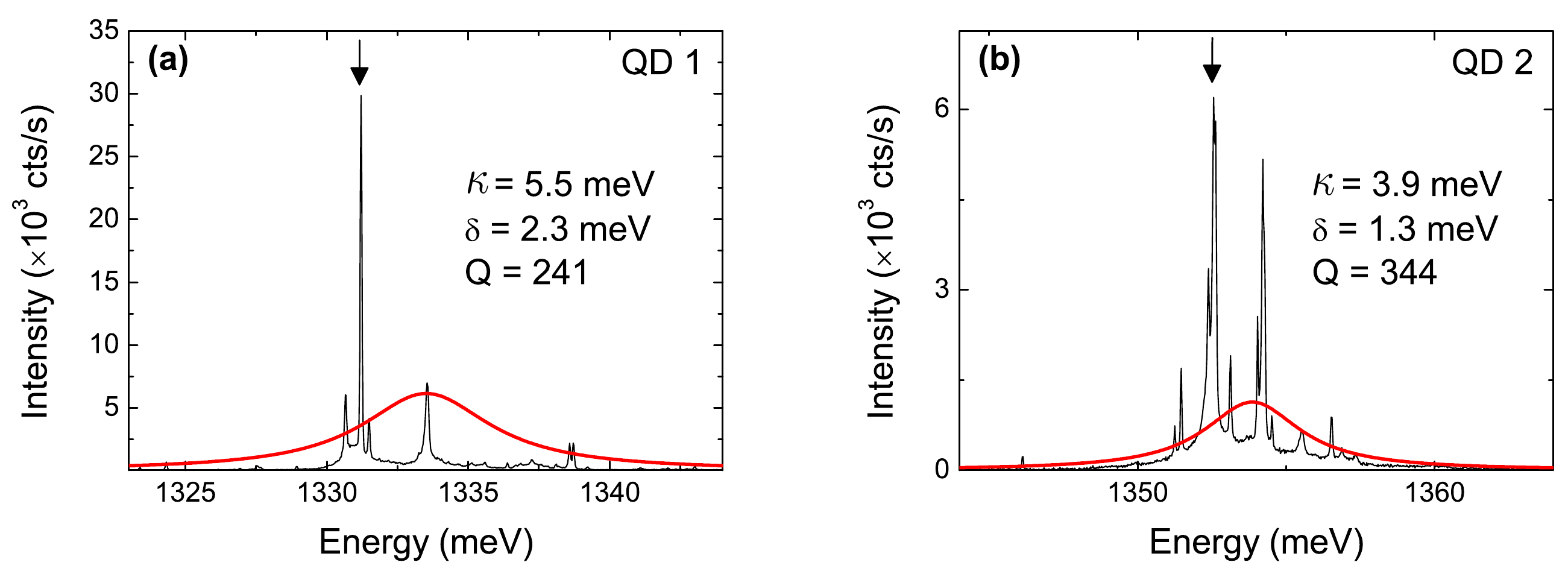}
\end{center}
\caption{High power non resonant excitation spectra for QD1 and QD2 at $4 \, \mathrm{K}$. The red line is the transfer function of the cavity and is extracted by fitting the background luminescence (removing the QDs luminescence). The single exciton transition used in each sample is indicated in the figure by the arrow.}
\label{cavity_mode}
\end{figure}

InAs/GaAs self-assembled QDs were grown by molecular beam epitaxy (MBE) on a planar (001) GaAs substrate and embedded in a planar microcavity made of unbalanced Bragg mirrors, with 24 pairs below and 12
pairs above the QDs. To create single-mode one-dimensional waveguides, ridges ranging from $1 \, \mu \mathrm{m}$ to $3 \, \mu \mathrm{m}$ wide, depending on the sample, were etched approximatively $1.5 \, \mu \mathrm{m}$ deep by inductively coupled plasma etching. Passivation of the surface has  further  improved  the  quality  of  the  heterointerfaces, leading to an important suppression of the scattered laser and allowing the realization of resonant excitation~\cite{melet2008resonants, PhysRevB.90.041303s}. Indeed, in this geometry, the excitation laser propagates along the 1D waveguide, while RF is collected in a perpendicular direction as shown in Fig.~1a of the main text. The cavity has little effect as a spectral filter but rather enhances the QDs RF collection by a factor of 20 to 50. The quality factor is rather low around $Q\sim200-300$ corresponding to width $\kappa\approx 4-5 \, \mathrm{meV}$. We show in Fig.~\ref{cavity_mode} high power non resonant excitation spectra for QD1 and QD2. We clearly see that only the QDs in the cavity mode can be observed and we extract the transfer function of the cavity by fitting the background luminescence (removing the QDs luminescence). We indicate by an arrow the luminescence lines of QD1 and QD2 on which TPI have been performed and we extract for each one the width of the cavity and the detuning between the QDs under study and the cavity mode. These values are used in the fitting procedure in Fig.~3 of the main text and in Fig.~\ref{V_fct_T_DQ2} in the Supplement.

\section{Fourier transform spectroscopy measurements}

\begin{figure}[h!]
\begin{center}
\includegraphics[scale=0.45]{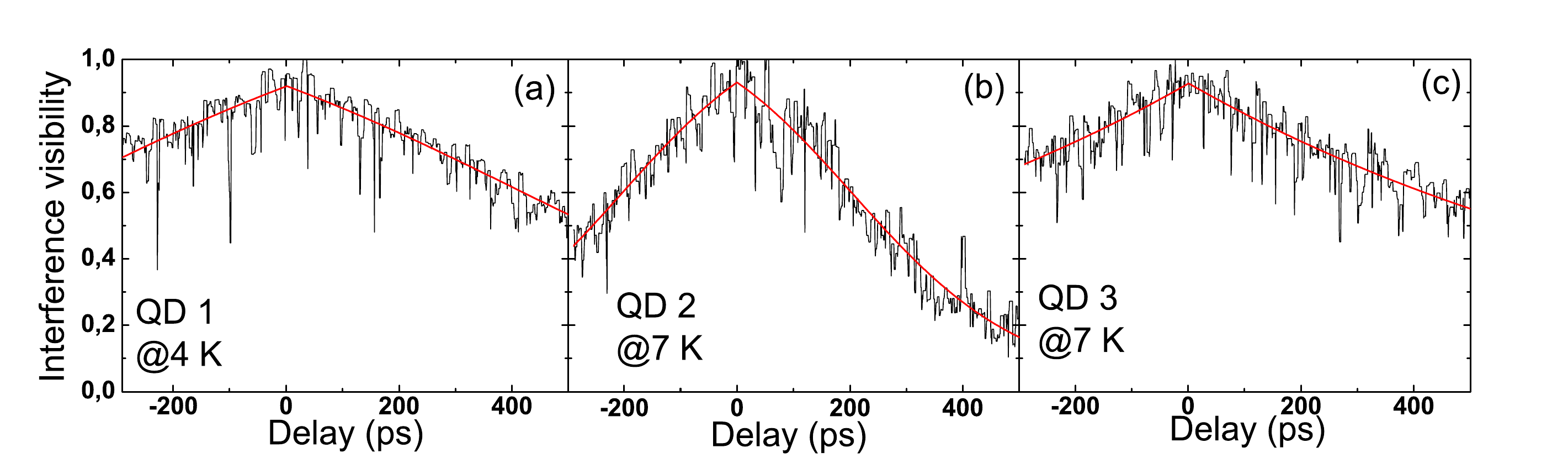}
\end{center}
\caption{Fourier transform spectra for the three studied QDs. The acquisition time for each position of the delay line is $1 \, \mathrm{s}$ which means that we average all decoherence processes that take place at shorter time scale. The red solid lines are fits using Eq.~\eqref{pseudo_voigt}.}
\label{T2_measurements}
\end{figure}

We determine the coherence time $T_2$ of the RF of the QDs from first-order correlation function measurements using a Michelson interferometer. As said in the main text, the contrast of the interference fringes is fitted by a pseudo-Voigt profile (see Eq.~\eqref{pseudo_voigt}) which allows to extract $T_2$ and the inhomogeneous contribution $\eta$.
\begin{eqnarray}
f(t) = (1 - \eta) \mathrm{e}^{-|t|/T_2} + \eta \mathrm{e}^{-(t/T_2)^2} \label{pseudo_voigt}
\end{eqnarray}
This inhomogeneous contribution gives rise to a Gaussian shape of the interference contrast and reflects the slight energy shift of the RF because of fluctuating charges trapping in the vicinity of the QD~\cite{kuhlmann2013charges, berthelot2006unconventionals}. Figure~\ref{T2_measurements} shows the results of Fourier transform spectroscopy (FTS) for the three QDs mentioned in the main text.
The errors on $T_2$ are of the order of $10-15 \, \%$ for QD 1 and 3 of the order of $5 \, \%$ for QD 2. They are mainly due to the path-length difference between the two arms of the interferometer that should be large compare to $c T_2$, $c$ being the light velocity. For the same reason, the errors on $\eta$ are of the order of 0.1.

\section{Second order correlation function and TPI visibility measurements}

\subsection{Second order correlation function for HBT experiment}

As said in the main text, the unusual shape of the histogram of coincidences for HBT experiment is due to the fact that the single photons pass through the two arms of the Mach-Zehnder interferometer used for the HOM setup (see Fig.~1a in the main text). For this experiment, the laser delivers single 3-ps pulses at 82MHz. The contribution to the coincidences histogram at 0 and $\pm3 \, \mathrm{ns}$ is attributed to the remaining scattered laser. Indeed, the peak labelled 2 (at zero delay) in Fig 2a of the main text, corresponds to coincidences of photons taking the same path of the Mach-Zenhder interferometer. The peaks labelled 1 and 3 (at $\pm3 \, \mathrm{ns}$ delay) are due to coincidences of photons taking each a different path of the Mach-Zenhder interferometer.

We measure $g^{(2)}_{\mathrm{HBT}}$ by normalizing the three central peaks integrated intensity by the averaged integrated intensity of the two three-peaks bunches at (12, 9, and $15 \, \mathrm{ns}$) and (-12, -9, and $-15 \, \mathrm{ns}$). The red solid-line corresponds to a multi-exponential decay fit which allows to take into account the overlapping of the different peaks.

\subsection{TPI visibility}
To extract a quantitative value of the second order correlation function at zero delay for HOM experiment $g^{(2)}_{\mathrm{HOM}}$, we follow a procedure close to the one describes in ref.~\cite{santori2002indistinguishables}. This method has the advantage to take into account the presence of the remaining scattered laser (obtained by HBT measurements) and the imperfections of the experimental setup. The coincidence histograms for HBT and HOM experiments are fitted by a multi exponential decay allowing to extract the value of $A_i^{\mathrm{HBT}}$ and $A_i^{\mathrm{HOM}}$ ($i \in \{1,2,3\}$) which are respectively the areas of the peak labeled $i$ in Figure~2a and b of the main text. In the case of an ideal experimental setup and for perfect single photon emitter, the value of the second order correlation function at zero delay for HOM experiment would be given by:
\begin{eqnarray}
g^{(2)}_{\mathrm{HOM}}(0) = \frac{A_2^{\mathrm{HOM}}}{A_1^{\mathrm{HOM}} + A_3^{\mathrm{HOM}}}
\end{eqnarray}

\begin{figure}[h!]
\begin{center}
\includegraphics[scale = 0.39]{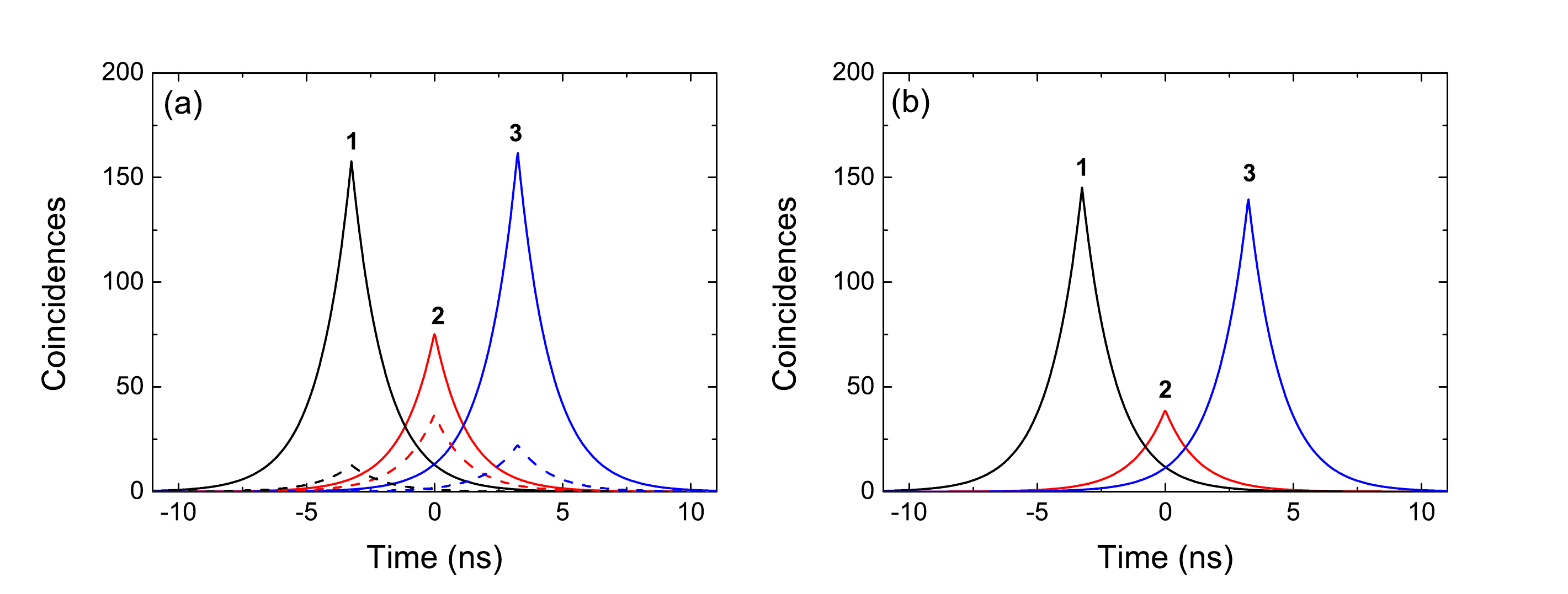}
\end{center}
\caption{(a) Fits for the three central peaks for HOM (solid line) and HBT (dashed line) experiments for QD~1 at $T = 4 \, \mathrm{K}$ presented in Fig.~2 of the main text. For the HBT contribution, the fits have been multiplied by the factor $2 t_{\mathrm{HOM}}/t_{\mathrm{HBT}}$. (b) Result of Eq.~\eqref{Eq_cor_HOM_SM} allowing to take into account the remaining laser background.}
\label{manip_HOM_HBT_SM}
\end{figure}

The first deviation to the ideal case describes above is due to the presence of the remaining scattered laser mixed to the luminescence of the QD and send into the Mach-Zehnder interferometer. The areas of the peak 1, 2 and 3 are then corrected as follow:
\begin{eqnarray}
B_i = A_i^{\mathrm{HOM}} - 2\frac{t_{\mathrm{HOM}}}{t_{\mathrm{HBT}}}A_i^{\mathrm{HBT}} \label{Eq_cor_HOM_SM}
\end{eqnarray}
where $t_{\mathrm{HBT}}$ and $t_{\mathrm{HOM}}$ are the acquistion times for HBT and HOM experiments respectively. This correction shown in Figure~\ref{manip_HOM_HBT_SM}a and b allows to remove from the HOM histogram the laser/QD coincidences. The factor 2 comes to the fact that for HOM experiment two pulses are sent into the interferometer. This procedure is equivalent to the presence of the terms proportional to $2g$ in equation~2 of ref.~\cite{santori2002indistinguishables} without assuming that the parameter g characterises the two-photon emission probability but only characterising the multi-photon emission probability. Note that these two procedures neglect the coincidences due to the laser between the two pulses.

The second deviation to the ideal case is due to the contrast of the Mach-Zenhder interferometer $C \neq 1$ and not perfect 50/50 fibered beam splitter (FBS). In our case we measured, $R = 0.430 \pm 0.005$, $T = 0.570 \pm 005$ and $C^2 = 0.98 \pm 0.02$. The value of the TPI visibility is then given by
\begin{eqnarray}
V_{\mathrm{TPI}} = \frac{R^2 + T^2}{2RTC^2} \big( 1 - g^{(2)}_{\mathrm{HOM}} \big)
\end{eqnarray}

\section{Effects of the presence of an He-Ne laser on the TPI visibility} \label{HeNe_SM}

\begin{figure}[h!]
\begin{center}
\includegraphics[scale = 0.39]{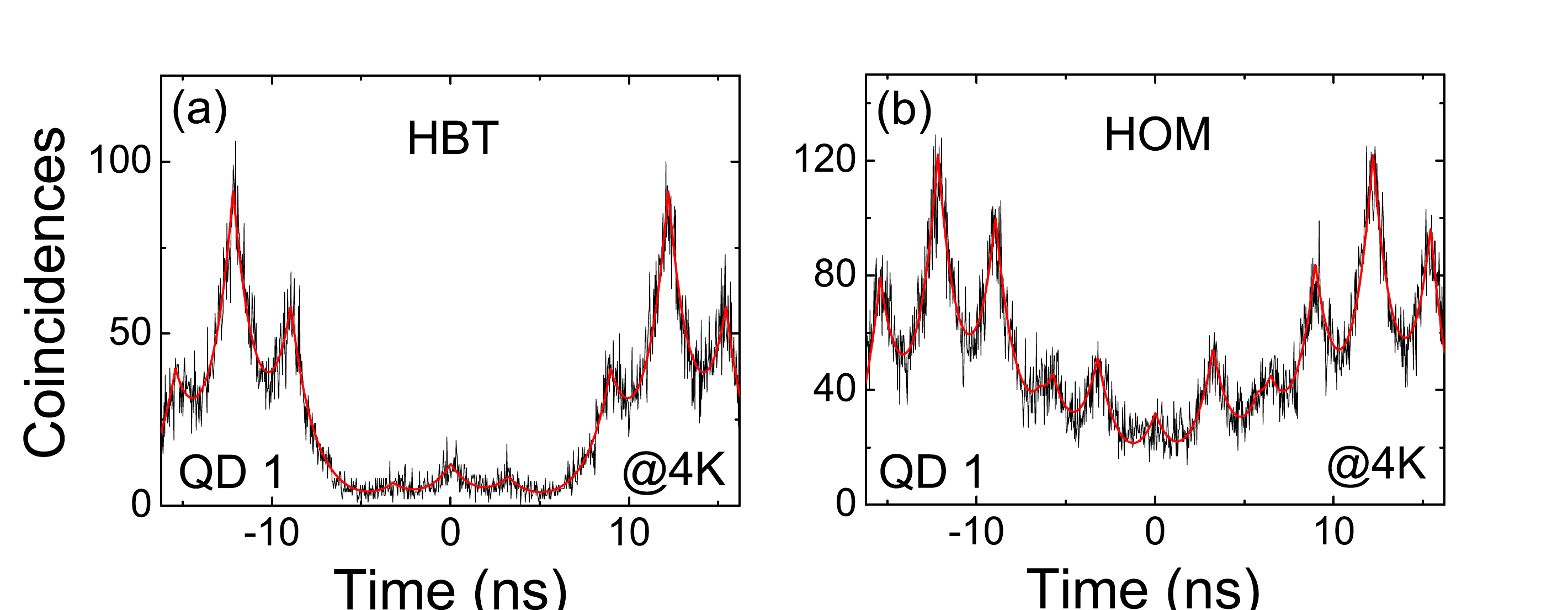}
\end{center}
\caption{Second-order correlation measurements for QD1 at $ 4 \, \mathrm{K}$ without He-Ne laser to enhance the RF for $t_{\mathrm{HBT}} = 8500 \, \mathrm{s}$ and $t_{\mathrm{HOM}} = 5200 \, \mathrm{s}$. (a) Coincidences histogram for HBT experiment, we extract $g^{(2)}_{\mathrm{HBT}} = 0.14 \pm 0.02$. (b) Coincidences histogram for TPI experiment. After correction by the remaining laser background, we obtain $V_{\mathrm{TPI}} = 0.80 \pm 0.05$.}
\label{manip_HBT_HOM_no_HeNe_SM}
\end{figure}

Let us notice that for QD1 we used a few nW of an additional He-Ne laser to enhance the RF while for QD2 and QD3 this non resonant laser allows to recover the RF~\cite{nguyen2012opticallys, PhysRevB.90.041303s}. We show here that this He-Ne laser has no influence on the measured TPI visibility, by presenting results on QD1 at $4 \, \mathrm{K}$ without the He-Ne laser (see Fig.~\ref{manip_HBT_HOM_no_HeNe_SM}). We measured $V_{\mathrm{TPI}} = 0.80 \pm 0.05$ whereas $V_{\mathrm{TPI}} = 0.79 \pm 0.03$ with a low power of He-Ne laser (see Fig.~2 of the main text).

\section{Quadratic electron-phonon coupling}

We start by consider the general electron-phonon coupling Hamiltonian $H_{e-ph} = H_0 + H_I$. Here $H_0 = \sum_m\omega_m\ket{m}\bra{m} + \sum_\vk\nu_\vk b^\dagger_\vk b_{\vk}$ is the energy of the uncoupled QD and phonon environment, where $\ket{m}$ is the $m^{th}$ electron-hole states of the QD and $\omega_m$ is its energy. 
The $\vk^{th}$-mode of the phonon environment is described by the creation and annihilation operators, $b_\vk^\dagger$ and $b_\vk$ respectively, with frequency $\nu_\vk$.
For linear electron-phonon coupling~\cite{mahan2013manys}, the interaction Hamiltonian is given by:
\begin{equation}
H_I = \sum\limits_\vk M_\vk\rho(\vk)(b_\vk^\dagger + b_\vk),
\end{equation}
where $M_\vk = {k v(\vk)/\sqrt{2\varrho\mathcal{V}\nu_\vk}}$ is the phonon matrix element, here $\varrho$ is the mass density of the material, $\mathcal{V}$ is the phonon quantisation volume, and $v(\vk)$ is the deformation potential energy of the phonons. 
We have also defined $\rho(\vk) = \int d^3 r e^{i\vk\cdot\vecr}\rho(\vecr)$ as the Fourier transform of the particle density operator $\rho(\vecr)  = \sum_{jj^\prime}\psi_j^\ast(\vecr)\psi_{j^\prime}(\vecr)c_{j}^\dagger c_{j^\prime}$, where $c_j^\dagger$ is the fermionic creation operator of the $j^{th}$-electronic state, and $\psi_j(\vecr)$ is its corresponding wavefunction.

This Hamiltonian couples all electronic states together through phonon transitions. 
To simplify our dynamical description of the system we follow Muljarov and Zimmerman~\cite{muljarov2004dephasings}. 
This involves removing the off-diagonal transitions to higher lying excited states in the QD through an effective Hamiltonian approach~\cite{cohen1992atoms}, restricting our description to a TLS system with ground state $\ket{0}$ and the single exciton state $\ket{X}$, with splitting $\omega_X$. This results in a Hamiltonian that is quadratic in the electron-phonon coupling~\cite{muljarov2004dephasings}:
\begin{equation}\label{HAM}
H = \omega_X\ket{X}\bra{X} + \ket{X}\bra{X} \left[\sum\limits_\vk g_\vk \left( b_\vk^\dagger + b_\vk\right) +\sum\limits_{\vk,\vk^\prime} f_{\vk,\vk^
\prime} \left( b_\vk^\dagger + b_\vk\right)\left( b_{\vk^\prime}^\dagger + b_{\vk^\prime}\right) \right] + \sum\limits_\vk \nu_\vk b_\vk^\dagger b_\vk.
\end{equation}
The coupling constants are given by:
\begin{equation}
g_\vk = \bra{X}M_\vk\varrho(\vk)\ket{X}\hspace{1cm}\text{and}\hspace{1cm} f_{\vk,\vk^\prime} = \sum\limits_{m>1}\frac{\bra{X}M_\vk\varrho(\vk)\ket{m}\bra{m}M_{\vk^\prime}\varrho(\vk^\prime)\ket{X}}{\omega_X-\omega_m},
\end{equation}
where $f_{\vk,\vk^\prime}$ contains a sum over the states $\ket{m}$ describing the electronic configurations of the QD.
We can gain analytic expressions for these coupling constants by assuming the electron/holes in the QD are in an isotropic harmonic potential, and the QD has mirror symmetry (i.e. the electron and hole experience the same confinement potential). 
The linear coupling takes the standard form $g_\vk = (2\varrho\mathcal{V}c_s)^{-1/2}\sqrt{k}(D_e-D_h)e^{-c_s^2k^2/2\nu_c^2}$, where we have assumed linear dispersion for the phonons with speed of sound $c_s$, and we have introduced the cut-off frequency of the phonon environment $\nu_c$.

For the quadratic coupling, we consider only virtual transition between the lowest excited state and the first manifold of next excited states. 
This yields:
\begin{equation}
f_{\vk,\vk^\prime} =- \sum\limits_{j\in\{x,y,z\}}\frac{c_s}{2\varrho\mathcal{V}\nu_c^2}\left(\frac{D_e^2}{\Delta_e} +\frac{D_h^2}{\Delta_h} \right) \sqrt{k k^\prime}k_jk^\prime_j e^{-\frac{c_s^2(k^2 + k^{\prime 2})}{2\nu_c^2}},
\end{equation}
where $k_j$ are the Cartesian components of the wavevector.

\section{Polaron transformation and master equation}

To derive an equation of motion for the above Hamiltonian, we make use of the polaron transformation, $\mathcal{U}_p = \outter{X} e^{S} + \outter{0}$, where $S = \sum_k\nu_\vk^{-1}g_\vk(b_k^\dagger - b_k)$~\cite{nazir2016modellings}. 
Applying this to the Hamiltonian in Eq.~\ref{HAM}, we obtain:
\begin{equation}\begin{split}
H_V &= \left(\omega_X - \sum\limits_k\frac{g_\vk^2}{\nu_\vk}\right)\ket{X}\bra{X}+\sum\limits_\vk \nu_\vk b_\vk^\dagger b_\vk + \ket{X}\bra{X}\sum\limits_{\vk\vk^\prime}f_{\vk,\vk^\prime}(b_\vk^\dagger + b_\vk)(b_{\vk^\prime}^\dagger + b_{\vk^\prime}),
\end{split}\end{equation}
where any residual linear coupling is zero due to spherical symmetry. 
We now wish to use this Hamiltonian to derive a second-order master equation for the density operator of the QD, $\chi(t)$, which simply takes a pure dephasing form:
\begin{equation}\label{MEQ}
\frac{\partial\chi(t)}{\partial t} = -i\left[\tilde\omega_x\ket{X}\bra{X},\chi(t)\right] +\Gamma\mathcal{L}_{\sigma}[\chi(t)]+ 2\gamma_{pd}\mathcal{L}_{\sigma^\dagger\sigma}[\chi(t)],
\end{equation}
where $\tilde\omega_X =\omega_X - \sum_\vk g^2_\vk/\nu_\vk$ is the polaron shifted resonance of the TLS, and $\mathcal{L}_O[\chi] = O \chi O^\dagger - \left\{O^\dagger O,\chi\right\}/2$.
Notice that have introduced a second dissipator to account for the optical transition in the QD with lifetime $\Gamma$.
The pure dephasing rate is then given by $\gamma_{pd} = \operatorname{Re}\left[\sum_{\vk,\vk^\prime}\int_0^\infty d\tau\vert f_{\vk,\vk^\prime}\vert^2\langle\mathcal{B}_\vk(\tau)\mathcal{B}_{\vk^\prime}(\tau)\mathcal{B}_\vk(0)\mathcal{B}_{\vk^\prime}(0)\rangle\right]$, where $\mathcal{B}_\vk(\tau) = b_\vk^\dagger e^{ic_s k\tau} + b_\vk e^{-ic_s k\tau}$.
We can make a technical simplification by assuming that the dominant contribution to the virtual phonon scattering occurs due to phonons scattering into different modes, that is only scattering processes where $\vk\neq\vk^\prime$.
This allows us to factorise the $\vk$ and $\vk^\prime$ components such that:
\begin{equation}
\gamma_{pd} =\frac{\pi}{c_s} \frac{\mathcal{V}^2}{(2\pi)^6}\int d^3 k\int d^3 k^\prime\vert f(\vk,\vk^\prime)\vert^2\left\{n(k^\prime)(n(k) + 1)\delta(k-k^\prime) + n(k)(n(k^\prime) + 1)\delta(k-k^\prime) \right\},
\end{equation}
where we have taken the continuum limit and used the definition of the $\delta$-function, $\delta(x) =\pi^{-1}\operatorname{Re}[ \int_0^\infty d\tau e^{ix\tau}]$. 
We have also defined the bosonic occupation number $n(k) = [\exp(-\beta\nu_k)-1]^{-1}$, where $\beta^{-1} = k_B T$ is the thermodynamic temperature. 
Resolving the $\delta$-function, and using spherical symmetry we obtain:
\begin{equation}
\gamma_{pd} 
=
\frac{\alpha^2\mu}{\nu_c^4}\int\limits_0^\infty  \nu^{10} e^{-2\nu^2/\nu_c^2}n(\nu)(n(\nu)+1)d\nu ,
\end{equation}
where we have assumed linear dispersion to write the above in terms of frequency, and we have defined the constants $\alpha = [4\pi^2\varrho c_s^5]^{-1}(D_e - D_h)^2$ and $\mu = \pi[D_e - D_h]^{-4} ( \Delta_e^{-1}D_e^2 +\Delta_h^{-1}D_h^2)^2$.

\section{Indistinguishability in the Polaron frame}

We now wish to calculate the indistinguishability/visibility of two photon interference. 
To do so, we use the definition of the indistinguishability in frequency space~\cite{ilessmithFuture}:
\begin{equation}
\mathcal{I} = P^{-2}\int\limits_{-\infty}^\infty d\omega\int\limits_{-\infty}^\infty d\nu \vert\mathcal{S}(\omega,\nu)\vert^2,
\end{equation}
where $\mathcal{S}(\omega,\nu) = \langle  \hat{E}^\dagger(\omega) \hat{E}(\nu)\rangle = \int_0^\infty dt_1\int_0^\infty dt_2 g^{(1)}(t_1,t_2) e^{-i\omega t_1} e^{i\nu t_2}$ is the generalised two-colour spectrum, 
$g^{(1)}(t_1,t_2) = \langle\hat{E}^\dagger(t_1)\hat{E}(t_2)\rangle$ is the first-order correlation function, and $P = \int_{-\infty}^{\infty}\langle \hat{E}^\dagger(\omega)\hat{E}(\omega)\rangle d\omega$ is the total light emitted from the QD. 
$\hat{E}(t_1)$ and $\hat{E}(\omega)$ are the Heisenberg picture field operators written in the time- and frequency-domain respectively.

In order to relate the detected field operators to the emission of the QD, we must account for emission via the sideband and spectral filtering due to the cavity. 
Though the low-Q cavity is much broader than the ZPL, its presence will lead to some filtering of the phonon sideband. 
In frequency space the field after the filter takes the form $\hat{E}(\omega)  = h(\omega) \hat{E}_{QD}(\omega)$, where $h(\omega) = (\kappa/2)[i(\omega - \delta) +(\kappa/2)]^{-1}$ with $\kappa$ defined as the cavity width, $\delta$ the cavity-QD detuning, and where $\hat{E}_{QD}(\omega)$ is the unfiltered field operator of the QD.
In the time domain the QD field in the polaron frame is given by $\hat{E}_{QD}(t) = \sqrt{\Gamma/2\pi}\sigma(t)B_-(t)$, where $B_{\pm}  = e^{\pm S}$ is the displacement operator of the phonon environment~\cite{iles2016fundamentals,ilessmithFutures}.
With this definition the first order correlation function for the QD becomes $g^{(1)}(t,\tau) = 2\pi \Gamma\langle B_+(\tau)B_-\rangle\langle\sigma^\dagger(t+\tau)\sigma(t)\rangle$, where the first term describes the phonon sideband, and the second the zero phonon line contribution to the spectra~\cite{ilessmithFutures}.
Using the quantum regression theorem, in addition to the master equation given in Eq.~\ref{MEQ}, we can write an analytic form for the 
first order correlation function $g^{(1)}(t,\tau) = 2\pi \Gamma B^2 \mathcal{G}(\tau)\exp(-\Gamma t -(\Gamma+2\gamma_{pd})\tau/2)$, 
where $\mathcal{G}(\tau) = \exp(\varphi(\tau))$ is the phonon correlation function, $B = \exp(-\varphi(0)/2)$ is the Franck-Condon factor, 
and $\varphi(\tau) = \int_0^\infty d\nu \nu^{-2}J(\nu)\left(\coth\left(\nu/2k_BT\right)\cos(\nu\tau) - i\sin(\nu\tau)\right)$.
We have introduced the spectral density $J(\nu)=\sum_\vk\vert g_\vk\vert^2\delta(\nu-\nu_\vk) = \alpha\nu^3\exp(-\nu^2/\nu_c^2)$.

By recognising that phonon processes occur on a time scale much faster than optical transitions, 
we can separate the phononic and photonic contributions~\cite{ilessmithFutures} in the generalised two colour 
spectrum such that $\mathcal{S}(\omega,\nu) = \mathcal{S}_{ZPL}(\omega,\nu) + \mathcal{S}_{SB}(\omega,\nu)$, 
where $\mathcal{S}_{ZPL}(\omega,\nu) = B^2 h(\omega)^* h(\nu) \Gamma(i(\omega-\nu)+\Gamma + 
2\gamma )\left[\left(\frac{1}{2}(\Gamma + 2\gamma )-i\nu\right)(\Gamma -i(\nu-\omega))\left(\frac{1}{2}(\Gamma + 2\gamma )+i\omega\right)\right]^{-1} $, 
which reduces to a (Lorentzian filtered) Lorentzian line shape when $\omega=\nu$. 
The term associated with the sideband is given by 
$\mathcal{S}_{SB}(\omega,\nu) = S_{SB}(\omega,\nu)+S^\ast_{SB}(\nu,\omega)$, 
with
\begin{equation}
S_{SB}(\omega,\nu) \approx B^2 h^\ast(\omega)h(\nu)\int\limits_0^\infty dt e^{i(\nu-\omega)t}g^{(1)}(t,0)\int\limits_0^\infty d\tau(\mathcal{G}(\tau) - 1)e^{-i\omega\tau} = \frac{\Gamma B^2 h^\ast(\omega)h(\nu)}{\Gamma - i(\nu-\omega)} S_{PH}(\omega)
\end{equation}
where we have defined the sideband spectrum $S_{PH} (\omega)=\int_0^\infty d\tau(\mathcal{G}(\tau) - 1)e^{-i\omega\tau}$.

The phonon sideband is purely incoherent, thus it does not contribute to the numerator in the indistinguishability. 
Furthermore, making the approximation that the cavity is flat across relevant ZPL features, we can set $h(\omega)\approx h(0)$, 
and we therefore obtain $\mathcal{I}\approx 4\pi^2 |h(0)|^4 P^{-2}B^4\Gamma [\Gamma + 2\gamma_{pd}]^{-1}$. 
Using again that the cavity is much broader than the ZPL, we approximate $P\approx |h(0)|^2 P_{\mathrm{ZPL}} + \mathcal{F}P_{\mathrm{SB}}$, where 
$P_{\mathrm{ZPL}} = \int_{-\infty}^\infty S_{\mathrm{ZPL}}(\omega,\omega)d\omega= 2\pi B^2$ is the power in the ZPL, 
and $P_{\mathrm{SB}} =2\pi(1-B^2)$
is the power in the phonon sideband in the absence of a filter, while $\mathcal{F} = P_{\mathrm{SB}}^{-1} \int_{-\infty}^\infty \mathcal{S}_{\mathrm{SB}}(\omega,\omega)d\omega$ 
is the fraction of the phonon sideband not removed by the filter or cavity.
Putting this together, we obtain:
\begin{equation}
\mathcal{I} = \frac{\Gamma}{\Gamma+ 2\gamma_{pd}}\left(\frac{|h(0)|^2B^2}{|h(0)|^2B^2 + \mathcal{F}[1-B^2]}\right)^2,
\label{IWithSB}
\end{equation}
which is Eq.~(2) in the main text. 

For the fitting procedure, we can simplify this expression further by assuming weak electron--phonon coupling. 
Here, the phonon correlation function $\mathcal{G}(\tau) =\exp(\varphi(\tau)) \approx 1-\varphi(\tau)$, where we have considered only terms first-order in $\alpha$. 
The fraction of sideband unfiltered is given by $\mathcal{F} = 2 B^2P_{SB}^{-1}\int_{-\infty}^\infty d\omega \vert h(\omega)\vert^2\re[S_{\mathrm{PH}}(\omega)]$.
We can now carry out the Fourier transform of the phonon correlation function analytically, such that 
\begin{equation}\begin{split}
&\re[S_\mathrm{PH}(\omega)]\approx  \re\left[\int\limits_0^\infty\varphi(\tau)e^{-i\omega\tau}d\tau\right] 
= \pi\alpha\omega e^{-\omega^2/\nu_c^2}\left(\coth\left(\frac{\beta\nu}{2}\right)-1\right),
  \end{split}\end{equation}
where we have invoked the definition of the $\delta$-function. Using this expression, the filtered fraction becomes:
\begin{equation}
\mathcal{F} = P^{-1}_\mathrm{SB}\int_{-\infty}^\infty  \vert h(\omega)\vert^2 \omega e^{-\omega^2/\nu_c^2}\coth\left(\frac{\beta\omega}{2}\right)d\omega.
\end{equation}

\begin{figure}[t]
\begin{center}
\includegraphics[scale = 0.65]{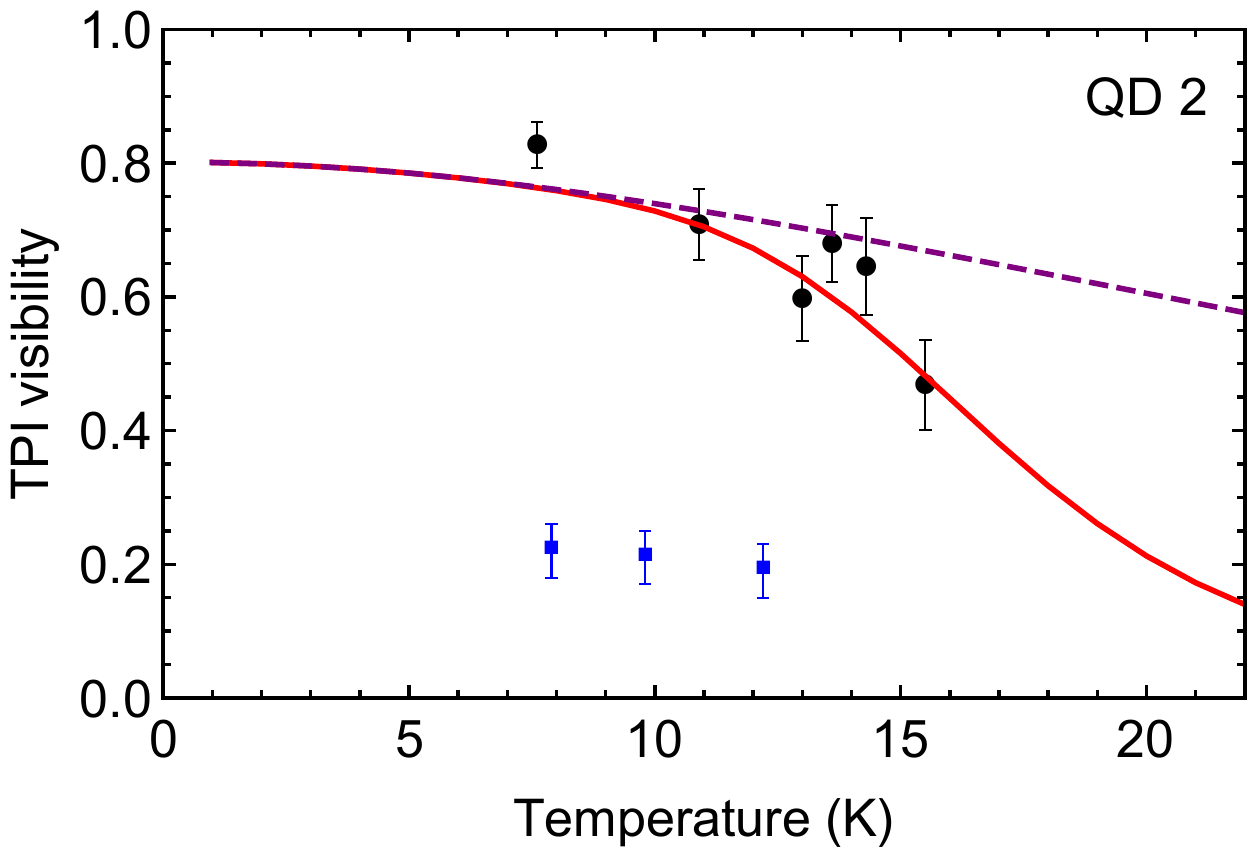}
\end{center}
\caption{Plot of the measured TPI visibility for QD2 as a function of temperature (black stars). The data are fitted with Eq.~2 of the main text. The full expression is used to produced the solid red curve and the dash purple curve shows only the effect of the phonon sideband on the TPI visibility.
The phonon parameters extracted from the fitting procedure are $\alpha = 0.0071 \, \mathrm{ps}^2$, $\nu_c = 11.9 \, \mathrm{ps}^{-1}$, $\mu = 5.6 \times 10^{-4} \mathrm{ps}^2$.}
\label{V_fct_T_DQ2}
\end{figure}

\section{TPI visibility for QD2}

We present in Fig.~\ref{V_fct_T_DQ2} temperature dependent measurement of the TPI visibility for QD2. These experiments were realized in another cryostat with a minimum temperature of $7$ K which explain the lack of data at low temperature. As for QD1 we use Eq.~2 of the main text to fit the experimental data with the same fitting procedure. Again, we observe very good agreement between the experimental data and the microscopic model. We extract the following parameters for QD2: $\alpha = 0.0071 \, \mathrm{ps}^2$, $\nu_c = 11.9 \, \mathrm{ps}^{-1}$, $\mu = 5.6 \times 10^{-4} \mathrm{ps}^2$ and $\mathcal{F} \sim 0.1$ at $T\sim7$~K. We notice the importance of taking into account virtual transitions to higher lying QD states in order to fit the rapid loss of the visibility for temperature higher than $10 \, \mathrm{K}$.

The blue squares represent the expected TPI visibility using the measured values of $T_1$ and $T_2$ and random dephasing processes by phonons and charges~\cite{bylander2003interferences}. We clearly see that these values do not reproduce the observed behaviour of the TPI visibility, confirming that TPI and FTS using a Michelson interferometer do not probe the same dephasing processes.

%





\end{document}